\documentclass[aps,prb]{revtex4}
\usepackage{graphicx,amsmath,amssymb,color,float}

\newcommand{\mc}{\mathcal}
\newcommand{\tet}{\texttt}
\newcommand{\ve}{\varepsilon}

\begin{document}
 \title{Temperature-dependent Plasmons and Their Damping Rates for Graphene with a Finite Energy Bandgap}
 \author{Andrii Iurov$^{1}$, Godfrey Gumbs$^{2,3}$,  Danhong Huang$^{4}$, and V. Silkin$^3$ }
\affiliation{$^{1}$ Center for High Technology Materials, University of New Mexico, Albuquerque, NM 87106 \\
$^{2}$ Department of Physics and Astronomy, Hunter College of the City University of New York, 695 Park Avenue, New York, NY 10065 \\
$^{3}$ Donostia International Physics Center (DIPC), P de Manuel Lardizabal, 4, 20018, San Sebastian, Basque Country, Spain \\
$^{4}$ Air Force Research Laboratory, Space Vehicles Directorate, Kirtland Air Force Base, NM 87117}
\date{\today}

\begin{abstract}
We obtained numerical and closed-form analytic expressions for finite-temperature plasmon dispersion
relations for intrinsic graphene in the presence of a finite energy gap in the energy 
spectrum. The calculations were carried out using the  random-phase approximation. The  analytic 
results have been derived in the high temperature regime and long-wavelength limit. We have found 
that the plasmon damping rate decreases in the presence of a band  gap. Our method of calculation
could also be applied to silicene and other buckled honeycomb lattice structures. The finite-temperature
plasmon dispersion relations are presented when a single graphene layer is Coulomb coupled to a semi
infinite conductor. Both cases of gapless  and gapped monolayer graphene  have been investigated when a
thick substrate is in their proximity. Both the plasmon excitation frequency and damping rate are linear
functions of the in-plane  wave vector in the long wavelength limit when a monolayer interacts with a 
conducting substrate which is not the case for free-standing pristine or gapped graphene.
\end{abstract}
\maketitle
\section{Introduction}
\label{sec1}
Plasmons in grpahene represent one of the most interesting and actively studied fields, from  both fundamentally theoretical
\cite{Wunsch, DS07, pavlo, SDSLi, chakraborty, GumbsBook, Rafael} and experimental or technological points of view \cite{1+, Poli1, Poli2}.
Graphene plasmons are especially important, partially because of their  versatile frequency range which may be adjusted
 by varying the doping concentrations of its mobile carriers and energy band gap.  Consequently,  graphene has several
potential  device applications in optics, microscopy, nanolithography \cite{GG1, GG2, GG3}. These  studies in
the area of plasmonics also extend to carbon-based structures such as fullerenes \cite{Nord1, Image, our1, our2, Anto, Liu2015},
carbon nanotubes \cite{nanotubes1Lin, nanotubes2}, as well as the recently discovered silicon-based silicene and other
buckled honeycomb lattice structures \cite{Silicene, silicene-canada} with  interesting roles played by  the on-site
potential differences between the sublattices. The main feature of the silicene energy dispersions is the energy gap, determined by the
spin-orbit interaction. We believe that our theoretical formalism, here applied  to gapped graphene, could  also
applicable in a straightforward way to silicene and germanene.

\medskip
\par
One of our goals is to investigate the way in which  the  energy band gap modifies the temperature-induced plasmon
excitations and their damping rates in intrinsic (or undoped) graphene. Specifically, we distinguish the
cases when the graphene layer is free-standing   and when it is in close proximity with a thick conducting substrate.
Clearly, such plasmon modes cannot be excited at $T=0$ because of the absence of free carriers \cite{SDSLi}. The gaplessness of graphene is an
important factor in the thermal population of the valence and conduction bands with electrons and hole correspondingly. Since the finite
temperature also results in the decay into particle-hole pairs, we are also interested to study how this decay (damping rate) depends on
the temperature.

\medskip
\par
Closed-form analytic expressions for the long wavelength plasmon dispersion relation in gapped
graphene as well as the dynamical polarization  function  were  at $T=0$ K initially reported by
 Pyatkovskiy in Ref.[\onlinecite{pavlo}]. This important study demonstrated  the existence of an
extended range of wave vectors for the undapmed plasmons in the presence of a finite energy gap.
This bandgap could be opened using a substrate or by exposing the graphene layer to  circularly-polarized light.
\cite{Kibis} The polarization function in this case was obtained analytically in Ref.[\onlinecite{Busl}].

\medskip
\par
We also consider finite-temperature plasmons in graphene  which is Coulomb-coupled to a semi-infinite conducting
substrate. For possible tailoring of the plasmon frequencies, graphene has been combined with prefabricated
plasmonic nanoarrays and metamaterials in order to obtain hybrid plasmon devices \cite{PR1, PR2, PR3}. Therefore,
a thorough understanding of the  dispersion and damping of plasmons in graphene  interfacing with
different kinds of substrates is necessary for  producing innovative practical applications. We have investigated
zero-temperature non-local plasmons in one of  our previous studies \cite{Arx3}, applying the theoretical formalism
for a graphene layer interacting with a surface plasmon in a semi-infinite  conductor was developed in \cite{Horing, NJMH}.
The plasma instability in such systems was addressed in Ref.[\onlinecite{Arx2}]. These results  could be considered as
a non-trivial extension of Ref.[\onlinecite{DasSarma}]  in which a linear acoustic plasmon mode was obtained for two
interacting graphene layers.

\par
\medskip
\par
Some of our work is devoted to a careful calculation of the analytical results for the real and imaginary
parts of the polarization function in the long wavelength limit and for the dispersion equation which
 yields the plasmon modes. The rest of the paper is organized as follows:
in Sec.\ref{sec2} we show that the finite-temperature polarization function used in Ref.[\onlinecite{SDSLi}]
may be extended to the case of gapped graphene. After that, in Sec.\ref{sec3}, we employ  these results to
gapped graphene to obtain the high-temperature plasmon dispersion relation in the long-wave limit, as well as the
corresponding damping rate. This is an important contribution
of our study. In Sec.\ref{sec4} we derive the finite-temperature plasma dispersion relation for a graphene layer  which is Coulomb-coupled
to a semi-infinite substrate. These analytical results show  novel non-trivial behavior, connected with the temperature dependence of
each plasmon dispersion  for both real and imaginary parts of the plasmon frequency. Finally,  numerical results
for the plasmon dispersion using the full polarization function  for gapped graphene
are presented in Sec.\ref{sec5}.

\section{High-Temperature Plasmon Dispersion Relation for Gapped Graphene    }
\label{sec3}

We now consider gapped graphene with energy dispersion: $\varepsilon(k) = \pm \sqrt{\Delta^2+(\hbar v_F k)^2}$
where $v_F$ is the fermi velocity and $\Delta$ is the energy gap between the valence (-) and conduction (+) bands
and the polarization function in the long wavelength limit is given by

\begin{equation}
\label{gap1}
   \Pi^{0}(q,\omega; \Delta) = \frac{2\mu}{\pi \hbar^2}\left(1-\left( \frac{\Delta}{\mu}\right)^2 \right) \frac{q^2}{\omega^2}
\end{equation}
Equation\ {(\ref{gap1})} is only valid  for $\Delta < \mu$. If $\Delta > \mu$, the valence band is completely empty
and only inter-band transitions contribute to the plasmon excitations. The polarization functions is then
given by

\begin{equation}
 P_0 (q,\omega; \Delta) = \frac{-2 }{\pi \hbar} \frac{q^2}{v_F^2 q^2 - \omega^2} \left\{
 2\Delta + \frac{\hbar^2 \left( v_F^2 q^2 - \omega^2  \right) - 4 \Delta^2}{\sqrt{v_F^2 q^2 - \omega^2}} \
 \arcsin  \left(
 \frac{v_F^2 q^2 - \omega^2}{v^2 q^2 - \omega^2 - 4 \Delta^2/\hbar^2}
\right)^{1/2}
\right\} \ .
\end{equation}
For $\Delta \to  0$,  we have

\begin{equation}
P^{0} (q,\omega; \Delta \rightarrow 0) = \frac{-2 g}{\pi \hbar} \frac{q^2}{v_F^2 q^2 - \omega^2} \left\{
0 + \sqrt{v_F^2 q^2 - \omega^2} \arcsin{1}
\right\} = \frac{- i g}{4} \frac{q^2}{\sqrt{\omega^2 -v_F^2  q^2}}
\end{equation}
which is similar to Eq.\ (\ref{main1}).

\par
We  now turn our attention to evaluate the finite-temperature polarization

\begin{eqnarray}
&&	\Pi^{(0)}_T(q,\omega;\Delta) = \int\limits_{0}^{\infty} d\,\mu' \frac{\Pi^{0}(q,\omega)}{4 k_BT
	\left( \cosh \left[ \frac{\mu - \mu'}{2 k T}
		\right] \right)^2} = \\
\nonumber				
&&	=	\int\limits_{0}^{\mu' = \Delta} d\mu^\prime\
		\frac{P_{0}(q,\omega; \Delta)}{4 k_BT
		\left( \cosh \left[ \frac{\mu - \mu'}{2 k_BT}
		\right] \right)^2}+\int\limits_{\mu' = \Delta}^{\infty} d\,\mu' \
		\frac{2\mu'}{\pi \hbar^2}\left(1-\left( \frac{\Delta}{\mu'}\right)^2 \right)
		 \frac{q^2/ \omega^2}{4 k_BT \left( \cosh \left[ \frac{\mu - \mu'}{2 k_BT}
		\right] \right)^2}
\end{eqnarray}
This expressions may be expressed as a sum of three parts, i.e.,
$ \Pi^{(0)}_T(q,\omega;\Delta) = \sum\limits_{j=1}^{3}I_j$ with

\begin{eqnarray}
  I_1 &=& \int\limits_{0}^{\mu' = \Delta} \frac{P_{0}(q,\omega; \Delta)}{4 k_BT \left( \cosh \left[ \frac{\mu - \mu'}{2 k_BT}
		\right] \right)^2} = P_{0}(q,\omega; \Delta) \int\limits_{0}^{\mu' = \Delta} \frac{1}{4 k_BT \left( \cosh \left[ \frac{\mu - \mu'}{2 k_BT}
		\right] \right)^2} = \frac{1}{2} P_{0}(q,\omega; \Delta)\tanh\left[\frac{\Delta}{2 k_BT}  \right]
\nonumber\\
I_2 &=&\int\limits_{\mu' = \Delta}^{\infty} d\,\mu' \frac{2\mu'}{\pi \hbar^2}
		 \frac{q^2/ \omega^2}{4 k_BT \left( \cosh \left[ \frac{\mu - \mu'}{2 k_BT}
		\right] \right)^2} = \frac{2 q^2}{\pi \hbar^2 \omega^2}
\left\{
		\frac{k_BT}{4} \ln 16 - \frac{\Delta}{2} \tanh\frac{\Delta}{2 k_BT} + k_BT \ln \left[ \cosh \frac{\Delta}{2 k_BT} \right]
\right\}		
\nonumber\\
 I_3 &=& \frac{\Delta^2 q^2}{2 k_BT \pi \hbar^2 \omega^2}
 \int\limits_{\Delta}^{\infty} \frac{d\, \mu'}{\mu'} \left( \cosh[(\mu - \mu')/(2 k_BT)]\right)^{-2}
\end{eqnarray}
which are still to be evaluated.

\subsection{Plasmon Dispersion relation (${\mbox Re}\  \, \omega_p$)}

\begin{equation}
P_0(q,\omega;\Delta) = \frac{1}{2 \pi \hbar^2} \frac{q^2}{v_F^2 q^2 - \omega^2} \left\{
 \frac{2 \Delta}{\hbar} + \frac{\pi}{2} \sqrt{v_F^2 q^2 - \omega^2}
 \right\}
\end{equation}
which consists of two terms, namely,

\begin{equation}
  P_0(q,\omega;\Delta) = - \frac{i}{4 \hbar} \frac{q^2}{\sqrt{\omega^2 - v_F^2 q^2}} -
	\frac{q^2}{\pi \hbar^2} \frac{\Delta}{\omega^2 - v_F^2 q^2} \ .
\end{equation}
Therefore,  for $T=0$ K, the first correction which is associated with the energy gap
is linear, real and negative. We now collect all the approximations from each term and write

\begin{equation}
 I_1 = \frac{1}{2} P_0(q,\omega; \Delta) \tanh\left[ \frac{\Delta}{2 k_BT} \right] \backsimeq -
\frac{1}{\pi \hbar^2} \frac{\Delta^2}{4 k_BT}
 \frac{q^2}{\omega^2} \ .
\end{equation}
We note that the zero-order term is purely imaginary.
We present the next part  as

\begin{equation}
 I_2 = \frac{1}{\pi \hbar^2} \frac{k_BT}{2} \frac{q^2}{\omega^2} \ln 16
+ \frac{2}{\pi \hbar^2} \frac{q^2}{\omega^2} G_1(T,\Delta) \, ,
\end{equation}
where

\begin{equation}
 G_1(T,\Delta) = k_BT \ln\left[ \cosh \frac{\Delta}{2 k_BT} \right] -
\frac{\Delta}{2} \tanh\frac{\Delta}{2 k_BT} \backsimeq
 \frac{\Delta^2}{8 k_BT} - \frac{\Delta^2}{4 k_BT} = - \frac{\Delta^2}{8 k_BT}  \  .
\end{equation}
In summary,  we have

\begin{equation}
 I_2 = \frac{2 \ln 2}{\pi \hbar^2}  k_BT \frac{q^2}{\omega^2} - \frac{1}{\pi \hbar^2} \frac{\Delta^2}{4 k_BT} \frac{q^2}{\omega^2}
\end{equation}
The  remaining term is defined by

\begin{equation}
 I_3 = \frac{1}{2 \pi \hbar^2} \frac{\Delta^2}{2 k_BT} \frac{q^2}{\omega^2} G_2(T, \Delta) \, ,
\end{equation}
where $G_2(T, \Delta)$ is a dimensionless  integral defined as

\begin{equation}
 G_2(T, \Delta) =  \int\limits_{\Delta}^{\infty} \frac{d\, \mu'}{\mu'}
\frac{1}{\left( \cosh[(\mu - \mu')/(2 k_BT)]\right)^{2}}  \ .
\end{equation}
The finite-temperature polarization function  may be expressed as

\begin{equation}
 \Pi^{(0)}_T (q,\omega; \Delta) =  \frac{2 \ln 2}{\pi \hbar^2}  k_BT \frac{q^2}{\omega^2} -
 \frac{1}{\pi \hbar^2} \frac{\Delta^2}{4 k_BT} \frac{q^2}{\omega^2} \left( 1 -  G_2(T, \Delta) \right)
\end{equation}
so that the  dielectric function $\epsilon (q, \omega)$ is

\begin{equation}
\epsilon (q, \omega) = 1 - \frac{2 \pi e^2}{\epsilon_s q} \Pi^{(0)}_T (q, \omega) = 1 -
 \frac{2 \pi}{q}  r_s \hbar v_F  \Pi^{(0)}_T (q, \omega) \ .
\end{equation}
From this, we deduce the temperature-induced plasma frequency of gapped graphene given by

\begin{equation}
\omega^2 = \frac{4}{\hbar} v_F  \, r_S \, q \, \left[k_BT \ln 2 -
\frac{\Delta^2}{4 k_BT} \left(1 - G_2(T, \Delta) \right) \right]
\label{GG1}
\end{equation}
\subsection{Evaluation of Integral  $G_2$}

Here,  we are looking for an approximated result in analytic form  for the following integral:

\begin{equation}
 \int\limits_{\delta \ll 1}^{\infty} \frac{dx}{x \cosh^2 x}  \ ,
\end{equation}
where $\delta=\Delta/k_BT$.     In order to avoid the singularity appearing when
$\delta \to 0$, we perform the integration by parts. Consequently, we obtain

\begin{equation}
 \int\limits_{\delta}^{\infty} \frac{dx}{x \cosh^2 x} =  \frac{\ln x}{\cosh^2 x} \Bigg|_{\delta}^{\infty} +
 2 \int\limits_{\delta}^{\infty} dx\   \frac{\ln x \tanh x \, }{\cosh^2 x}  \ .
\end{equation}
We estimate each term when $\delta \to 0$ or $\Delta \ll k_BT$. We have

\begin{equation}
 \frac{\ln x}{\cosh^2 x} \Bigg|_{\delta}^{\infty} = - \ln \left( \frac{\Delta}{2 k_BT}  \right)
\end{equation}
The second part is  a small correction since the integrand does not diverge for $\delta \to 0$, using

\begin{equation}
 \int\limits_{\delta}^{\infty} dx\ \frac{\ln x \tanh x \, }{\cosh^2 x} \backsimeq \int\limits_{0}^{\infty}  dx\
 \frac{\ln x \tanh x \,  }{\cosh^2 x} = \frac{1}{6} \left( 36 \ln \mc{A}_{GK} - 7 \ln 2- 3 (1 + \gamma_{Eu}) \right)  \, ,
\end{equation}
where $\mc{A}_{GK} \backsimeq 1.2824$ is the  Glaisher-Kinkelin constant, defined as

\begin{equation}
 \mc{A}  = \left\{ \left( \prod_{s=1}^{\nu - 1} s^{s} \right) \nu^{-\nu^2/2-\nu/2-1/12} \tet{e}^{\nu^2/4}
 \right\}_{\nu \rightarrow \infty}
\end{equation}
and $\gamma_{Eu} = 0.5772$ is the Euler-Mascheroni constant, given by the following expression:

\begin{equation}
 \gamma_{Eu} = \left\{ \sum_{s = 1}^{\nu} \frac{1}{s} -\ln \nu \right\}_{\nu \rightarrow 0}
\end{equation}
Numerical integration gives

\begin{equation}
 \int\limits_{0}^{\infty} dx\ \frac{\ln x \tanh x \, }{\cosh^2 x} = -0.1048
\end{equation}
and for $\delta = \Delta / (2 k_BT) = 0.05$, we have

\begin{equation}
 \int\limits_{\delta}^{\infty} dx\ \frac{\ln x \tanh x \, }{\cosh^2 x} = -0.1004 \ .
\end{equation}
So, finally, we arrive at the result

\begin{equation}
  G_2 (T, \Delta) = \int\limits_{\delta \ll 1}^{\infty} \frac{dx}{x \cosh^2 x} \backsimeq - \ln\left( \frac{\Delta}{2 k_BT} \right)
\end{equation}
and the plasmon frequency in Eq.\ (\ref{GG1}) is now given by the equation

\begin{equation}
\omega^2 = \frac{4}{\hbar} v_F  \, r_S \, q \, \left[ k_BT \ln 2 -
\frac{\Delta^2}{4 k_BT} \left(1 + \ln\left( \frac{\Delta}{2 k_BT} \right) \right)  \right]  \ .
\label{GG2}
\end{equation}
\begin{figure}
\centering
\includegraphics[width=0.45\textwidth]{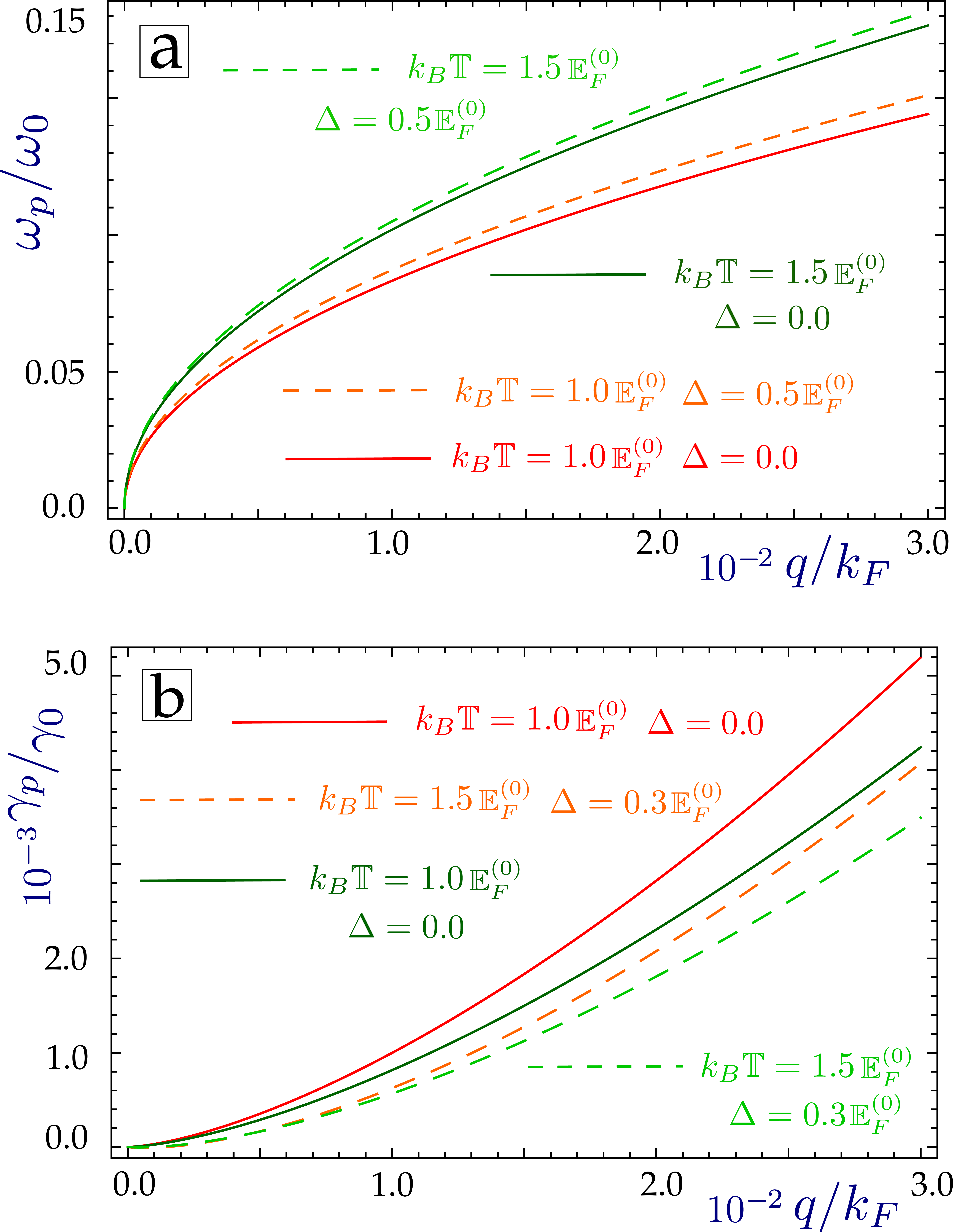}
\caption{(Color online) Plasmon frequency $(a)$ in units of
$\omega_0 = \left( 4/\hbar v_F r_s k_F \mathbb{E}_F^{(0)} \log2 \right)^{1/2}$, where $\mathbb{E}_F^{(0)}=\hbar v_F k_F  = 0.05\, eV$, and the damping rate $(b)$ in   units of
$\gamma_0 = \left( \pi^2/64 \,\, \hbar/\mathbb{E}_F^{(0)} \, r_s^{3} v_F^{3} k_F^{3} \log 2 \right)^{1/2}$
for   high-temperature plasmons in gapped graphene. Plasmon dispersions are shown for
$k_B \mathbb{T} = 1.0 \, \mathbb{E}_F^{(0)} \, \text{and} \, 1.5 \, \mathbb{E}_F^{(0)}$ and   energy
gap values $\Delta = 0.0 \, \text{and} \, 0.5 \mathbb{E}_F^{(0)}$, and   damping rates for
$\Delta = 0.0 \, \text{and} \, 0.3 \, \mathbb{E}_F^{(0)}$.}
\label{FIG:1}
\end{figure}
\subsection{ Plasmon Doping at a low temperature for extrinsic graphene }

\par
Let us consider the case of a finite doping $\mu > 0 $ and low temperature $T \to 0$. We must calculate:
the polarization function using

\begin{equation}
\nonumber
	\Pi^{(0)}_T(q,\omega;\Delta) = \int\limits_{0}^{\infty} d\,\mu' \frac{\Pi^{0}(q,\omega;\Delta)}{4 k_BT
	\left( \cosh \left[
	\frac{\mu - \mu'}{2 k_BT}
		\right] \right)^2} = \int\limits_{0}^{  \Delta} d\,\mu' \
		\frac{P_{0}(q,\omega; \Delta)}{4 k_BT \left( \cosh \left[ \frac{\mu - \mu'}{2 k_BT}
		\right] \right)^2}+\int\limits_{\mu' = \Delta}^{\infty} d\,\mu'
		 \frac{\Pi^{(0)}_{\mu'}(q, \omega, \Delta)}{4 k_BT \left( \cosh \left[ \frac{\mu - \mu'}{2 k_BT}
		\right] \right)^2} \, .
\end{equation}
The first term is trivial and given by

\begin{equation}
 \int\limits_{0}^{\mu' = \Delta} \frac{P_{0}(q,\omega; \Delta)}{4 k_BT \left( \cosh \left[ \frac{\mu - \mu'}{2 k_BT}
		\right] \right)^2} = P_{0}(q,\omega; \Delta) \frac{1}{2} \left\{
\tanh\left( \frac{\Delta - \mu}{2 k_BT} \right) + \tanh\left( \frac{\mu}{2 k_BT} \right)
\right\} \ .
\label{tanh}
\end{equation}
In the limit  $T \to 0$,  it follows that $\tanh(\alpha / T) \to \mbox{sign}(\alpha)$, so when
Eq.\ (\ref{tanh})  gives $2$  when $\Delta > \mu$  and $0$ when $\Delta < \mu$. So  that the integral
is $\theta(\Delta - \mu)$.
The second term is an integral representation of a Dirac delta-function for $T = 0$ K so that
$\delta(x) = \lim_{\epsilon \rightarrow 0} \frac{1}{2 \epsilon} \cosh^{-2} \left(x/
\epsilon  \right)$ and we have finally

\begin{equation}
\Pi^{0}(q,\omega;\Delta) = P_0(q,\omega;\Delta) + \theta(\mu - \Delta) \Pi_{\mu}^{0}(q,\omega;\Delta)
\label{GG3}
\end{equation}
as we had for $T=0$.
\begin{figure}
\centering
\includegraphics[width=0.6\textwidth]{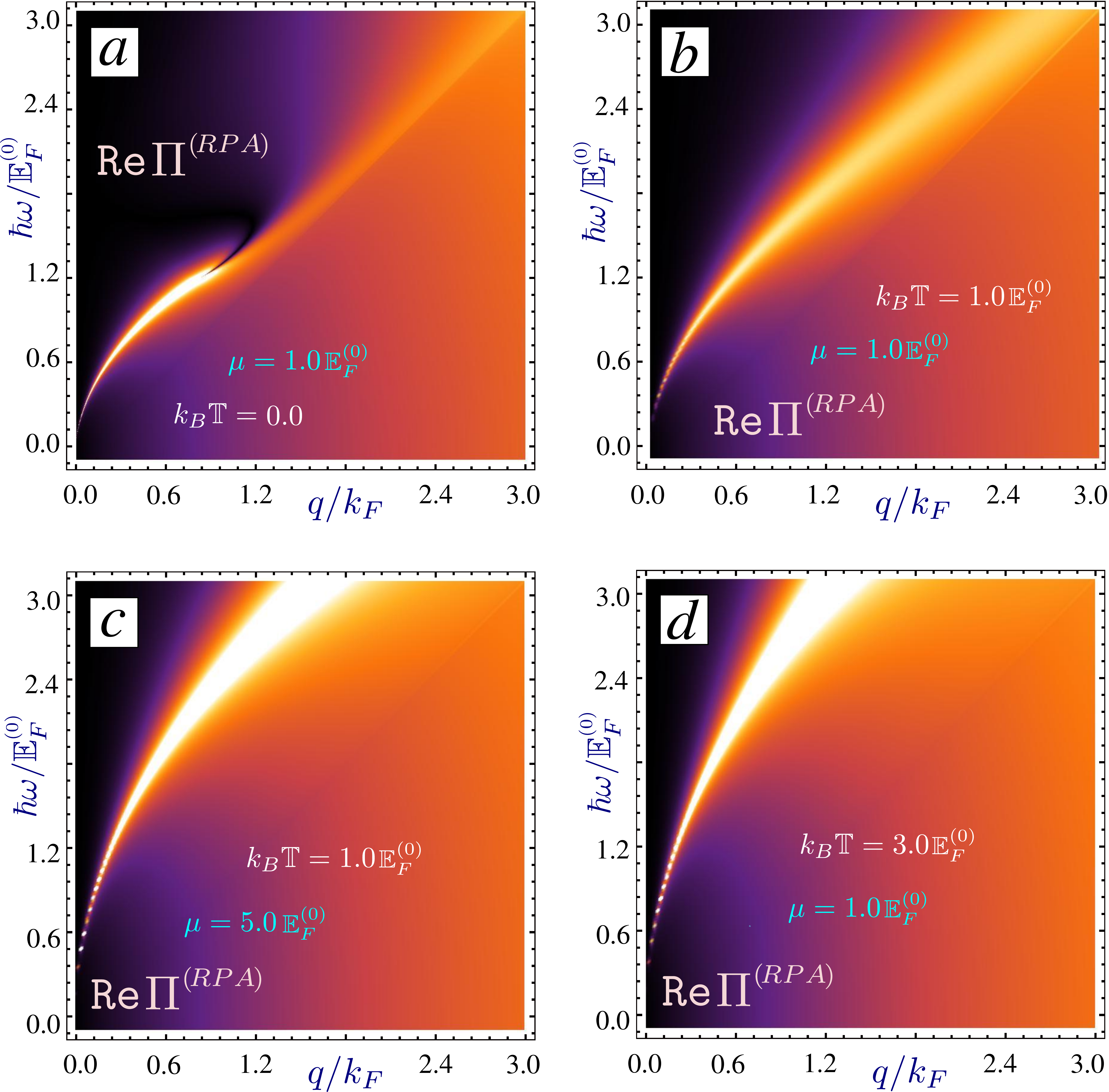}
\caption{(Color online) Finite-temperature plasmon excitations in graphene for various
chosen values of doping and temperature (density plot of RPA polarization function whose
peaks correspond  to the plasmon mode frequencies). Panel $(a)$ shows the zero-temperature
limit of a graphene plasmon with finite doping $\mu=1.0 \mathbb{E}_F^{(0)}$.
Plot $(b)$   corresponds to  $k_b \mathbb{T}=1.0 \mathbb{E}_F^{(9)}$ and $\mu=1.0 \mathbb{E}_F^{(0)}$.
Panels $(c)$ and $(d)$ show plots of  the plasmon dispersion relation for either high
temperature or doping value ($k_b \mathbb{T}=1.0 \mathbb{E}_F^{(0)}$ and $\mu=5.0 \mathbb{E}_F^{(0)}$
for $(c)$ and $k_b \mathbb{T}=3.0\mathbb{E}_F^{(0)}$ and $\mu=1.0 \mathbb{E}_F^{(0)}$ for $(d)$.}
\label{FIG:3}
\end{figure}
\begin{figure}
\centering
\includegraphics[width=0.6\textwidth]{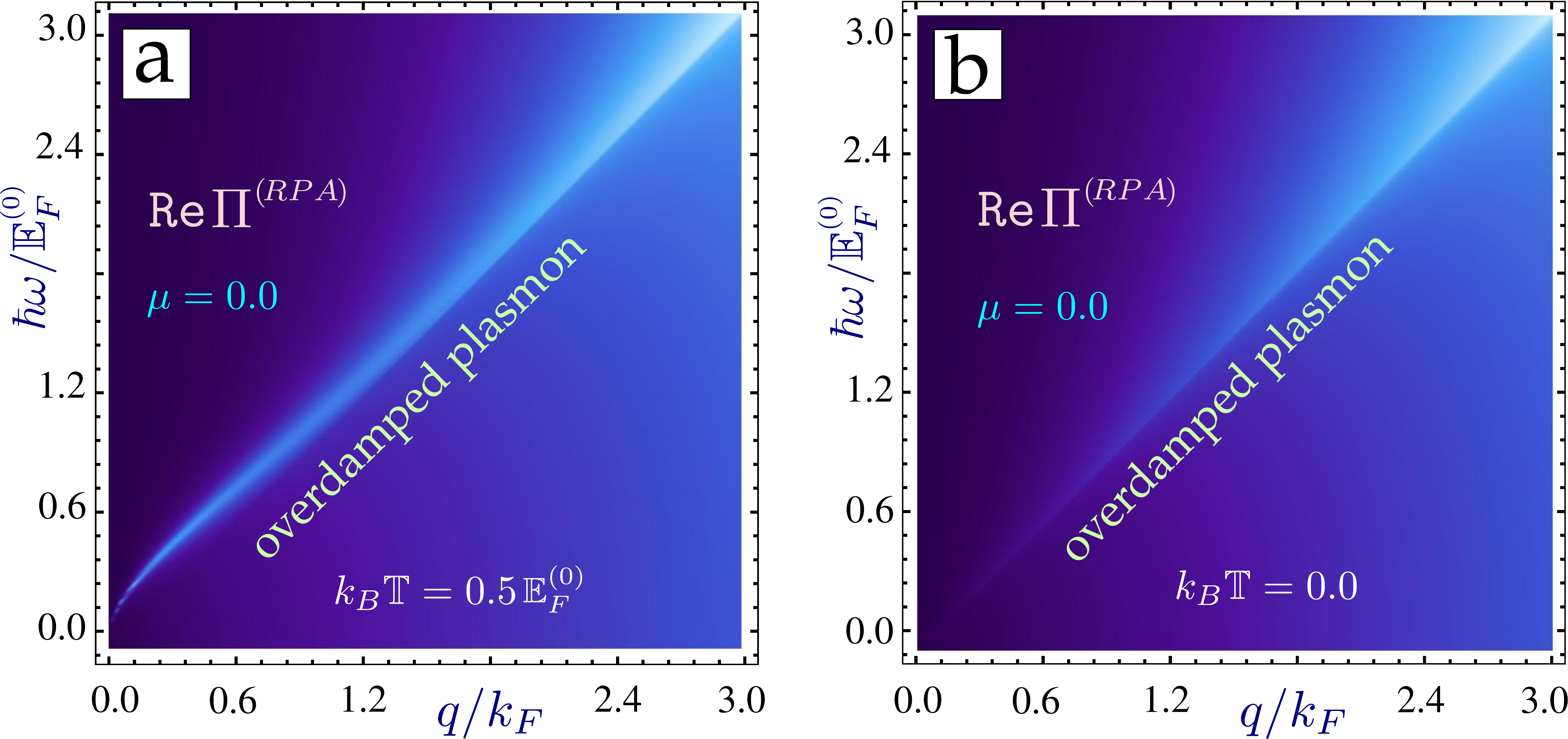}
\caption{(Color online) Overdamped (suppressed) plasmon frequencies for graphene at low doping
concentration and $T \to 0$. Panel $(a)$ gives the dispersion relation at low
 temperature $k_b \mathbb{T}=1.0 \mathbb{E}_F^{(0)}$, while panel $(b)$ corresponds to
  $\mathbb{T}=0$. Both panels are plotted for intrinsic monolayer graphene.}
\label{FIG:4}
\end{figure}
\subsection{Damping rates for gapped graphene}

The imaginary part of the zero-temperature polarization function  is from Eq. \ (\ref{GG3}) given by

\begin{equation}
{\mbox Im}\ \Pi^{0} (q, \omega) = \frac{g q^2}{8 \hbar \omega} \left\{
1- \frac{1}{2}X_0^2
\right\} \, \theta\left(2 \mu - \hbar \omega  \right) \, ,
\end{equation}
where

\begin{equation}
X_0 = \sqrt{1+ \frac{4 \Delta^2}{\hbar^2 \left( v_F^2 q^2 - \omega^2 \right)^2}} \backsimeq \left(
1 - \frac{2 \Delta^2}{\hbar^2 \omega^2}
\right)
\end{equation}
which gives

\begin{equation}
{\mbox Im}\ \Pi (q, \omega) = \frac{g q^2}{16 \hbar \omega} \left\{
1 + \frac{4 \Delta^2}{\hbar^2 \omega^2}
\right\} \, \theta\left(2 \mu - \hbar \omega  \right) \, .
\end{equation}
The finite-temperature polarizability is obtained as follows

\begin{equation}
{\mbox Im}\ \, \Pi^{(0)}_T (q, \omega) = - i \frac{g q^2}{16 \hbar \omega}  \left\{
1 + \frac{4 \Delta^2}{\hbar^2 \omega^2}
\right\} \int\limits_{0}^{\infty} \frac{d\,\mu'}{4 k_BT} \frac{\theta(\hbar \omega - 2 \mu' )}
{\cosh^2 \left[ \frac{\mu - \mu'}{2 k_BT} \right]}
\end{equation}
which has imaginary part

\begin{equation}
{\mbox Im}\ \, \Pi^{(0)}_T (q, \omega) = i \frac{g q^2}{16 \hbar \omega}  \left\{
1 + \frac{4 \Delta^2}{\hbar^2 \omega^2}
\right\} \frac{1}{2} \tanh \frac{\hbar \omega}{4 k_BT} \backsimeq  \frac{g}{64} \frac{q^2}{k_BT}  \left\{
1 + \frac{4 \Delta^2}{\hbar^2 \omega^2}
\right\} \ .
\end{equation}
Another correction to the imaginary part of the frequency

\begin{equation}
 {\mbox Im}\ \, P_0(q,\omega;\Delta) = - \frac{1}{4 \hbar} \frac{q^2}{\sqrt{\omega^2 - v_F^2 q^2}}
\end{equation}
This shows that when $T=0$ K, the first-order correction  associated with the energy gap
 is linear, real and negative. At finite temperature, our calculation shows that

\begin{equation}
 {\mbox Im}\ \, I_1 = \frac{1}{2} P_0(q,\omega; \Delta) \tanh\left[
\frac{\Delta}{2 k_BT} \right] \backsimeq - \frac{i}{4 \hbar} \frac{q^2}
{\sqrt{\omega^2 - v_F^2 q^2}} \frac{\Delta}{2 k_BT}
 \backsimeq - \frac{1}{8 \hbar} \frac{\Delta}{k_BT} \frac{q^2}{\omega} \ .
\end{equation}
Additionally, making use of  similar procedures when we dealt with zero gap, we obtain

\begin{equation}
- \frac{2 \gamma}{\omega^3}
\frac{q^2}{\pi \hbar^2} k_BT \left\{
2 \ln 2 - \frac{\Delta^2}{4 k^2 T^2} \left[
C + \ln \left( \frac{\Delta}{2 k_BT} \right)
\right]
\right\}
+ \frac{1}{16} \frac{q^2}{k_BT} \left\{
1 + \frac{4 \Delta^2}{\hbar^2 \omega^2} \right\}
- \frac{1}{8 \hbar} \frac{\Delta}{k_BT} \frac{q^2}{\omega} = 0
\end{equation}
from which we deduce the damping rate

\begin{equation}
\gamma = \frac{\omega^3}{32} \frac{\pi \hbar^2}{k^2 T^2} \left\{
1 - \frac{2 \Delta}{\hbar \omega} - \frac{2 \Delta^2}{\hbar^2 \omega^2}
\right\} /  \left\{
2 \ln 2 - \frac{\Delta^2}{4 k^2 T^2} \left[
C + \ln \left( \frac{\Delta}{2 k_BT} \right)
\right]
\right\}
\end{equation}
which may be rewritten as
\begin{equation}
\gamma = \frac{\pi}{8} \frac{\hbar^{1/2}}{\sqrt{k_BT}}
(\ln 2)^{1/2} \left( r_s v_F q \right)^{3/2} -
\frac{1}{8} \frac{r_s v_F q}{k_BT} \Delta - \left\{
\frac{2 \sqrt{k_BT} \left( q r_s v_F  \right)^{1/2}}{\hbar^{1/2} \sqrt{\ln 2}} -
\frac{(q v_F r_s)^{3/2}}{\hbar^{3/2} \sqrt{k_BT} \sqrt{\ln 2}} \ln \left[  \frac{\Delta}{2 k_BT} \right]
\right\} \Delta^2  \ .
\end{equation}
\begin{figure}
\centering
\includegraphics[width=0.6\textwidth]{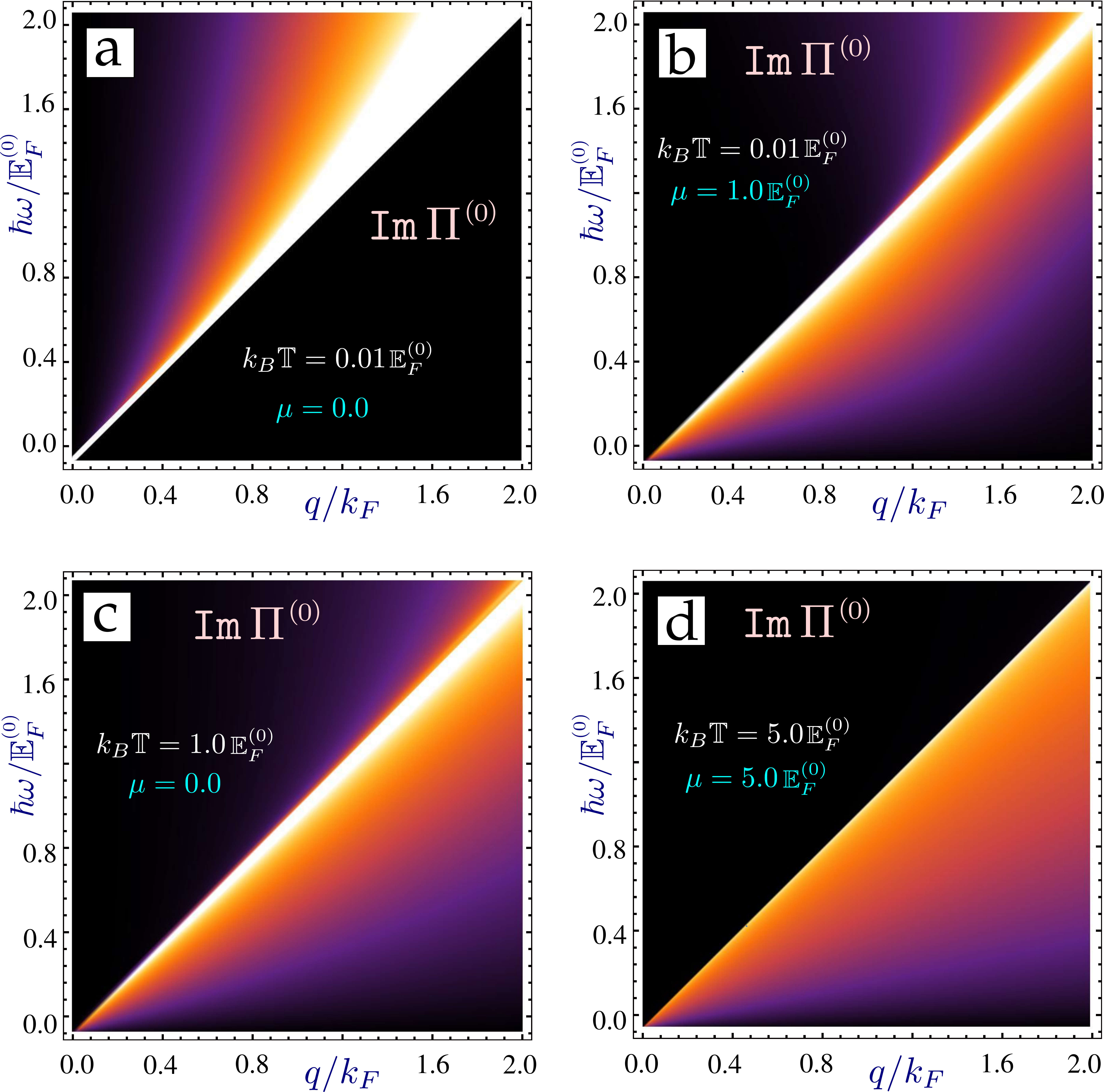}
\caption{(Color online)  Density plot of the imaginary part of the polarization function
 at various chosen  temperatures and doping concentrations in graphene. Panel $(a)$
 shows a plot corresponding to low   temperature $k_B \mathbb{T} = 0.01 \mathbb{E}_F^{(0)}$
and zero doping (in which plasmons do not exist due to the absence of free carriers).
Plot $(b)$ demonstrates the low-temperature limit for  finite $\mu = 1.0 \mathbb{E}_F^{(0)}$.
Panel $(c)$ corresponds to finite temperature and zero doping, whereas panel $(d)$ shows the
finite-temperature particle-hole modes for  highly doped graphene $\mu = 5.0 \mathbb{E}_F^{(0)}$.}
\label{FIG:2}
\end{figure}
\section{ Gapless Monolayer Graphene  Interacting with a Thick Conducting Substrate}
\label{sec4}

According to recent work\cite{Horing, Arx3}, the plasma dispersion relations  when a 2D layer is Coulomb-coupled
to a thick conducting substrate are determined by solving the following equation:

\begin{equation}
1 - v(q) \Pi^{(0)}_T (q,\omega) \left\{
1 + \tet{e}^{-2 a q} \frac{\omega_p^2}{2 \omega^2 - \omega_p^2 }
\right\} = 0 \ ,
\label{GG5}
\end{equation}
where $a$ is the layer-surface separation, $\omega_p$ is the bulk plasmon frequency and
the Fourier transform of the Coulomb potential energy

\begin{equation}
v(q) = \frac{2 \pi e^2}{\varepsilon_s q} = \frac{2 \pi}{q} \hbar r_s v_F
\end{equation}
is expressed in terms of $r_s$ is a dimensionless parameter.

\par
The finite-temperature polarization function is

\begin{equation}
\Pi^{(0)}_T (q,\omega) = \frac{2 \ln 2}{\pi} \frac{q^2}{\hbar^2 \omega^2} k_BT + \frac{i}{16} \frac{q^2}{k_BT}
\end{equation}
Here, we consider a well-defined plasmon with $\gamma \ll \omega$, which is excited with a certain
wave vector at a chosen temperature. Accordingly, we neglect the contributions to the real part
of the polarization function arising from the imaginary  part of the frequency.
So, making use of

\begin{equation}
{\mbox Re}\ \, \Pi^{(0)}_T (q,\omega) = \frac{2 \ln 2}{\pi} \frac{q^2}{\hbar^2 \omega^2} k_BT \, ,
\end{equation}
in Eq.\ (\ref{GG5}), we obtain

\begin{equation}
1 - \frac{2 \pi}{q} \hbar r_s v_F \frac{2 \ln 2}{ \, \pi} \frac{q^2}{\omega^2} k_BT
\left\{
1 + \tet{e}^{-2 a q}\frac{\omega_p^2}{2 \omega^2 - \omega_p^2}
\right\} = 0  \, .
\end{equation}
We rewrite this equation as

\begin{eqnarray}
&& 1 - \lambda \frac{q}{(\omega/\omega_p)^2} \left\{
1 + \tet{e}^{-2 a q}\frac{\omega_p^2}{2 \omega^2 - \omega_p^2}
\right\} = 0
\nonumber\\
&& \lambda = 4 \ln 2 \left(\frac{r_s v_F}{\hbar \omega_p^2} \right)\   k_BT \ .
\label{simp}
\end{eqnarray}
We now  consider \textit{two} different limiting cases. The first corresponds to when
the surface-layer separation is small so that $aq\ll 1$.  For this, we have the results
in the linear approximation

\begin{figure}
\centering
\includegraphics[width=0.6\textwidth]{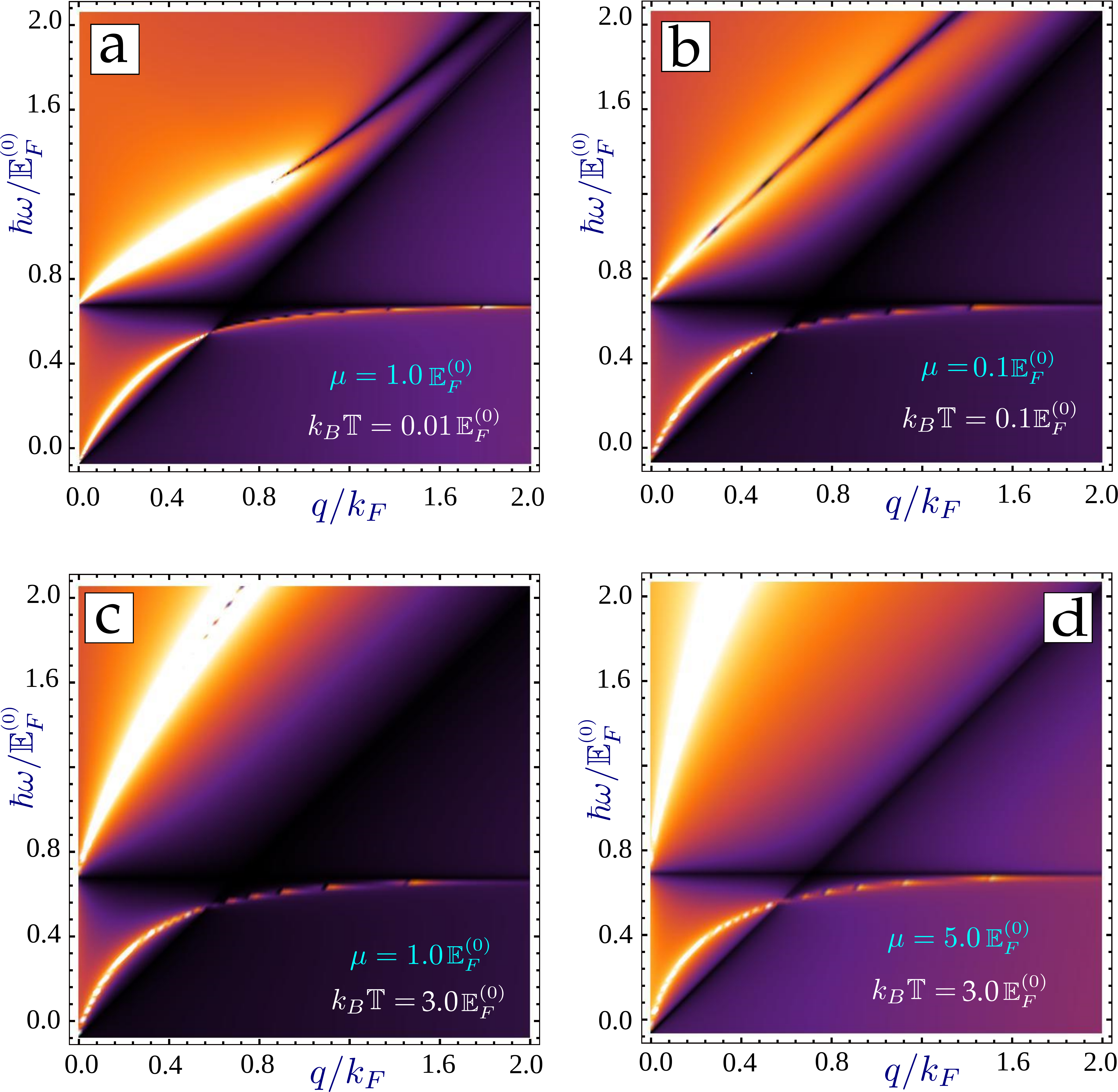}
\caption{(Color online) Plasmon dispersion relations for monolayer graphene, which is
Coulomb-coupled with a semi-infinite conductor for different chosen values of graphene
doping and temperature. Panel $(a)$ shows the low-temperature limit
when the doping is finite   corresponding to $\mu=1.0 \mathbb{E}_F^{(0)}$. Plot $(b)$
and shows the case for $k_b \mathbb{T}=0.1 \mathbb{E}_F^{(0)}$
for a slightly doped graphene  $\mu=0.1 \mathbb{E}_F^{(0)}$. Both panels $(c)$ and $(d)$
 correspond to a moderately high temperature $k_B \mathbb{T}=3.0 \mathbb{E}_F^{(0)}$ and
different chemical potentials ($\mu=1.0 \mathbb{E}_F^{(0)}$ for $(c)$ and
$\mu=5.0 \mathbb{E}_F^{(0)}$ for $(d)$).
 }
\label{FIG:5}
\end{figure}

\begin{eqnarray}
\Omega_1 &=& \sqrt{ 8 \ln 2 \frac{r_s v_F}{\hbar}} \sqrt{a}\sqrt{k_BT} \, q  \ .
\nonumber\\
\Omega_2 &=& \frac{\omega_p}{\sqrt{2}} + \sqrt{8} \ln 2 \, \frac{r_s v_F}{\hbar \, \omega_p} k_BT \, q
\end{eqnarray}
Both solutions are linear in $q$, we also note that the surface plasmon frequency does not depend on
$a$ until the second-order terms for the $\Omega_1$. branch.
Second-order terms in $q$ are neglected since we are interested in the temperature dependence
in the lowest order. Turning now to the set-up when $aq\gg 1$, a straightforward calculation
yields

 \begin{eqnarray}
\frac{\Omega_1}{ \omega_p}  &=& \sqrt{\lambda q} - \frac{1}{2}\frac{\sqrt{\lambda q}}
{1 - \lambda q} \tet{e}^{-2 a q}  \ ,
\nonumber\\
\frac{\Omega_2}{ \omega_p}  &=& \frac{1}{\sqrt{2}} + \frac{1}{\sqrt{2}} \frac{\lambda q}
{1 - \lambda q}  \tet{e}^{-2 a q} \ .
\end{eqnarray}
In the local  limit $\lambda q \ll 1$,  we obtain

\begin{eqnarray}
\Omega_1 &=& \sqrt{\frac{2 \ln 2}{\hbar}} \left( v_F r_s q \right)^{1/2} \sqrt{k_BT} \left\{
1 - \frac{1}{2} \tet{e}^{-2 a q}\right\}  \ ,
\nonumber\\
\Omega_2 &=& \frac{\omega_p}{\sqrt{2}} + 2 \sqrt{2} \ln 2 \, \frac{r_s v_F}{\hbar \, \omega_p} k_BT \, q \,
\tet{e}^{-2 a q} \ .
\end{eqnarray}

\medskip
\par

 We need to take into account the imaginary part of the $\omega_p^2/(2\omega^2 -   \omega_p^2)$
term. We make the replacement   $\omega \to \omega + i \gamma$ which leads to the  following
plasmon dispersion equation

\begin{equation}
1 - \left\{
\frac{4 \ln 2}{\hbar  } r_s v_F k_BT \frac{q}{\omega^2} \left(
1 - 2 i \frac{\gamma \omega_p^2 }{\omega^2}
\right)
 + i \frac{\pi}{8} r_s v_F \hbar \frac{1}{k_BT} q
\right\}
\left\{
1 + \tet{e}^{-2 a q} \left[ \frac{1}{2 (\omega/\omega_p)^2-1} - 4 i \gamma
\frac{\omega/\omega_p}{(2 (\omega/\omega_p)^2 - 1)^2}
\right]
\right\} = 0 \, ,
\end{equation}
From this, we obtain the   imaginary part for the lower ``acoustic'' branch to be

\begin{equation}
\Gamma_1 = \pi \sqrt{\frac{\ln 2}{2}} \sqrt{\frac{\hbar}{k_BT}} \left(r_s v_F\right)^{3/2} a^{3/2} q^3 \, ,
\end{equation}
while the ``upper'' branch has imaginary part

\begin{equation}
\Gamma_2 = \frac{\pi}{8 \sqrt{2}} \hbar v_F r_s \frac{\omega_p}{k_BT} q  \
\end{equation}
When the surface-layer separation  is large, the imaginary parts may be further approximated as

\begin{eqnarray}
\Gamma_1 &=& \frac{\pi}{8} \sqrt{\ln 2} \sqrt{\frac{\hbar}{k_BT}} (r_s v_F)^{3/2} \, q^{3/2}
\left[
1 - 8 \ln 2 \frac{r_s v_F}{\hbar \omega_p^2} k_BT\, q \,\tet{e}^{-2 a q}
\right] \, ,\, ,
\nonumber\\
\Gamma_2 &=&  \frac{\pi}{16} \hbar \omega_p \frac{r_s v_F}{k_BT} \, q \,\tet{e}^{-2 a q}
\end{eqnarray}
where $\lambda$ has been defined in Eq.\ (\ref{simp}).
We now investigate how  some of these results are affected when a finite energy gap exists
between the valence and conduction bands.

\section{Plasmons in Gapped Graphene, Coulomb-coupled to a Semi-infinite Conductor}
\label{sec5}

The real part of the polarization function may be expressed as

\begin{equation}
{\mbox Re}\  \, \Pi^{(0)}_T = \frac{2}{\pi \hbar^2} \frac{q^2}{\omega^2} \left\{
k_BT  \ln 2- \frac{\Delta^2}{8 k_BT}  \left[ C - \ln \left(
\frac{\Delta}{2 k_BT} \right) \right] \right\}  \, .
\end{equation}
Here $C = 1 - c_i \backsimeq 1 + 0.10 \backsimeq 1.10$.
Furthermore, the analytic form of Eq.\  (\ref{simp}) is still valid.
However, $\lambda$ must be replaced by $\lambda_\Delta$ defined as

\begin{equation}
\lambda_\Delta = 4 \, \frac{r_s v_F}{\hbar \, \omega_p^2}
\left\{
k_BT  \ln 2- \frac{\Delta^2}{8 k_BT}  \left[
C - \ln \left(
\frac{\Delta}{2 k_BT}
\right)
\right]
\right\}  \, .
\end{equation}

When $a q \ll 1 $,  the solutions are given approximately by

\begin{eqnarray}
\Omega_1 &=& \sqrt{8} \sqrt{a} \sqrt{\frac{r_s v_F}{\hbar}}
\left\{ k_BT  \ln 2- \frac{\Delta^2}{8 k_BT}  \left[
C - \ln \left( \frac{\Delta}{2 k_BT} \right)
\right]\right\}^{1/2} q
\nonumber\\
\Omega_2 &=& \frac{\omega_p}{\sqrt{2}} +  \sqrt{8} \, \frac{r_s v_F}{\hbar \, \omega_p}
\left\{ k_BT  \ln 2- \frac{\Delta^2}{8 k_BT}  \left[
C - \ln \left( \frac{\Delta}{2 k_BT} \right) \right]
\right\} \, q \ .
\end{eqnarray}

\medskip
\par 
on the other hand, when $ a q \gg 1$, the   solutions   are given simpler are given by

\begin{eqnarray}
\Omega_1 &=& \sqrt{\lambda_\Delta q} \left( 1 - 4 \tet{e}^{-2 a q} \right) = 2  \sqrt{\frac{r_s v_F}{\hbar}}
\left\{ k_BT  \ln 2- \frac{\Delta^2}{8 k_BT}  \left[
C - \ln \left( \frac{\Delta}{2 k_BT} \right)\right]
\right\}^{1/2}  \left( 1 - 4 \tet{e}^{-2 a q} \right) \, \sqrt{q}
\nonumber\\
\Omega_2 &=& \frac{\omega_p}{\sqrt{2}} + \sqrt{8} \omega_p \lambda_\Delta q \tet{e}^{-2 a q} =
\frac{\omega_p}{\sqrt{2}} +
8\sqrt{2} \, \frac{r_s v_F}{\hbar \, \omega_p}
\left\{ k_BT  \ln 2- \frac{\Delta^2}{8 k_BT}  \left[
C - \ln \left( \frac{\Delta}{2 k_BT}\right)\right]
\right\} q \tet{e}^{-2 a q} \, .
\end{eqnarray}
We now turn to presenting and discussing our numerical results in the next section.

\section{Numerical Results and Discussion}
\label{sec6}

The plasmon dispersion relations and damping rates for gapped graphene in the long-wavelength
 limit are presented in Fig.\ \ref{FIG:1}. The plasmon frequency   is increased at   higher
temperature, approximately having the behavior of $\backsimeq \sqrt{T}$ dependence as for
gapless graphene. Clearly, the damping rates  decrease with   increasing energy gap for all
temperatures. This means that the plasmon modes become less damped  in the presence of
a finite energy gap.

\medskip
\par
We are definitely interested in calculating the polarization function and plasmon energies
  numerically beyond the long wavelength limit. First, we obtain the imaginary
part of the polarization function in Fig.\ \ref{FIG:2}, looking for the regions within
$\{ q -\omega \}$-plane  where the relatively undamped plasmon modes exist $\left(
{\mbox Im}\  \Pi^{(0)}(q,\omega) \to 0 \right)$. As we know for the long wavelength limit,
 the imaginary part of the polarization function within the region of graphene plasmon
excitations decreases at high temperature as $\backsimeq 1/T$, \cite{SDSLi} so it is
relatively small at high temperatures although it not be negligible.  Panel $(b)$ of
Fig.\ \ref{FIG:2} demonstrates the behavior when$\mu = \mathbb{E}_F^{(0)}$ at very low
temperature, which approximately corresponds to the standard particle-hole modes of
zero-temperature graphene. \cite{Wunsch} We see that at  finite doping and  high temperature,
the region of  particle-hole modes (finite ${\mbox Im}\  \Pi^{(0)}(q,\omega)$) above the
diagonal $\omega = v_F q$ is suppressed, as shown in panel $(d)$.

\medskip
\par
Figure\ \ref{FIG:3} presents the  RPA polarizability  for various doping concentrations
 and temperatures. At zero temperature  and when the  doping is finite, we reproduce the
well-known  plasmon dispersion relation in graphene, as it was reported in Ref.\
[\onlinecite{Wunsch}]. As the temperature is increased, the plasmon
frequency is also  increased for all values of the wave vector, growing with temperature like
 $\sqrt{T}$ in the long wavelength limit as it was showed analytically in Ref.[\onlinecite{SDSLi}]).
Various examples of damped  plasmon excitations are shown in Fig.\ \ref{FIG:4}.
As it was mentioned above as well as in Ref.[\onlinecite{SDSLi}], the plasmon mode
has low intensity in the limit of vanishing temperature and zero doping.

\medskip
\par
The temperature-dependent plasmon dispersion  relation for a graphene  monolayer
interacting with a semi-infinite conductor is shown in Fig.\ \ref{FIG:5}. There are
clearly two plasmon branches, originating from the graphene layer (acoustic branch,
 starting at the origin), and the ``sruface plasmon'', which is depolarization shifted from
its long wavelength value of  at $1/\sqrt{2}\omega_p$. Both branches have positive
group velocity for $(q \to 0)$, i.e. are linear in the long-wave limit, similarly to its
variation at  zero temperature. Another interesting detail, which deserves mentioning
is that the higher plasmon branch, attributed to the surface behaves similarly to the
plasmons in graphene (see Fig.\ \ref{FIG:3}), while the lower  acoustic  branch remains
nearly unchanged for all the chosen temperature and doping values.

\medskip
\par
\begin{figure}
\centering
\includegraphics[width=0.45\textwidth]{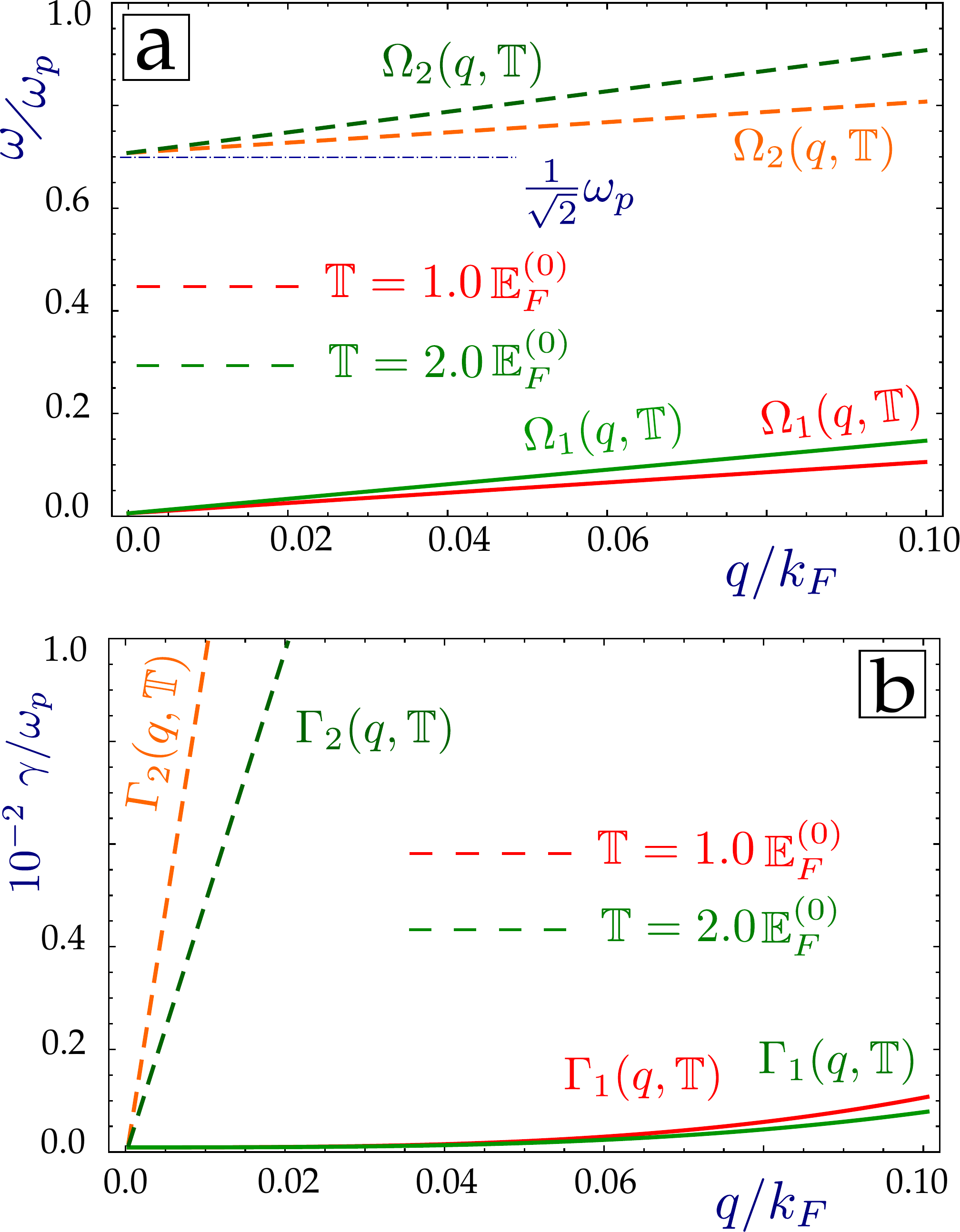}
\caption{(Color online)  $(a)$\ Plasmon frequencies  and  $(b)$ the damping rates  in the
 high-temperature limit for monolayer graphene interacting with  a semi-infinite conductor
in the long wavelength limit. Each plasmon branch (both real and imaginary parts)  are
 presented at two chosen temperatures -
$k_B \mathbb{T}=1.0 \mathbb{E}_F^{(0)}$ and $k_B \mathbb{T}=2.0 \mathbb{E}_F^{(0)}$.}
\label{FIG:6}
\end{figure}

The plasmon frequencies and the damping rates for grpahene, Coulomb-coupled with a thick
  conductor with free carriers in the long wavelength limit are presented in Fig.\ \ref{FIG:6}. Both
plasmon dispersions are linear and increase with  temperature as $\backsimeq \sqrt{T}$
and $\backsimeq T$, respectively. We also note that the damping rate $\Gamma_2$, which
corresponds to the upper plasmon, depends linearly on the wave vector $q$. This is a
new result, which was not encountered in graphene, either with a finite or zero energy gap.

\section{Concluding remarks}

In summary, we have obtained  analytic  expressions for the high-temperature plasmon dispersion
relations for  gapped graphene, as well as for a graphene layer interacting with a semi-infinite
conductor. We have found  that the plasmon frequency is modified according to
$\Delta^2/(4 k_BT) \ln ( \Delta/ (2 k_BT))$ in the presence of an  energy gap, which is
different from the case when $T=0$. The corresponding damping rate is decreased linearly
in the presence of a gap, making the plasmon more stable compared to its counterpart for
gapless graphene. Our investigation of a graphene monolayer, which is Coulomb-coupled to
 a semi-infinite metal, revealed a number of novel features for both  plasmon frequencies
and the damping rates. Specifically, we  emphasize the linear $q-$dependence of
some of the damping rates, which has not been encountered  for graphene (either gapped
 or ungapped).

\acknowledgments
This research was supported by  contract \# FA 9453-13-1-0291 of
AFRL.

\appendix

 \section{Finite temperature polarization - integral transformation derivation}
 \label{sec2}

In this Appendix, we prove that the finite-temperature polarizability could be expressed as an integral transformation
given in Eq. (\ref{DS1}) relating the polarization function $\Pi^{(0)}_T(q,\omega)$  at zero temperature, as
it was reported  in Ref.[\onlinecite{SDSLi}].

\par
We need to prove the general validity of the following expression:
	\begin{equation}
	\Pi^{(0)}_T(q,\omega) = \int  
	\limits_{0}^{\infty} d\mu^\prime\ \frac{\Pi^{0}(q,\omega)}{4 k_BT \left( \cosh \left[ \frac{\mu - \mu'}{2 k_BT}
		\right] \right)^2}
		\label{DS1}
	\end{equation}
with the zero-temperature polarization function is as follows:

\begin{equation}
\Pi^{0}(q,\omega) = \frac{g}{4 \pi^2} \int d^2 {\bf k} \ \
f^{s s'} ({\bf q},{\bf k})  \frac{n_F\left[ \varepsilon^s(k) \right] -
n_F\left[ \varepsilon^{s'}(\vert {\bf k} + {\bf q} \vert) \right]}{\varepsilon^s(k)
-  \varepsilon^{s'} (\vert {\bf k} + {\bf q} \vert)  + \hbar (\omega + i \gamma)}
\label{pi0}
\end{equation}
in terms of the spin and valley degeneracy factor $g=4$  and the form factor $f^{s s'} ({\bf q},{\bf k}) $.
 The only temperature-dependent terms in Eq.\ (\ref{pi0}) are the Fermi-Dirac
distribution functions $n_F (\varepsilon)= (1+ \tet{exp}[(\ve -\mu)/(k_BT)])^{-1} =
(1+ \tet{exp}[\beta (\ve -\mu)])^{-1}$ with $\beta = 1/(k_BT)$. This means that we must prove
a similar integral transformation for the   distribution functions.

\par
The Fermi-Dirac distribution functions could be presented as

\begin{equation}
n_F (\xi) = \frac{1}{2} \left( 1 - \tanh\frac{\beta \xi}{2}  \right) \, ,
\label{ff}
\end{equation}
where $\xi = \ve - \mu$. Indeed,
\begin{equation}
1 - \frac{2}{1+\tet{e}^{\beta \xi}} =  \frac{\tet{e}^{\beta \xi} - 1}{\tet{e}^{\beta \xi} + 1}
=  \frac{\tet{e}^{\beta \xi/2} - \tet{e}^{\beta \xi/2}}{\tet{e}^{\beta \xi/2}
+ \tet{e}^{\beta \xi/2}} = \tanh \frac{\beta \xi}{2}
\end{equation}
While at zero temperature, the distribution function could be presented as the Heaviside unit
 step with $n_F(\ve - \mu, T \rightarrow 0) = \theta(\mu - \ve)$. Accordingly,

\begin{equation}
\int\limits_{0}^{\infty} \frac{\theta(\mu - \ve) d \, \mu'}{4 k_BT \cosh^2 [(\mu - \mu')/(2k_BT)]}
= \int\limits_{\ve}^{\infty} \frac{d\, \mu'}{4 k_BT} \cosh^{-2} \left[\frac{\mu - \mu'}{2 k_BT}\right] \, .
\end{equation}
We substitute $\eta = \frac{\mu - \nu'}{2 k_BT}$ so that $d \, \mu' = - 2 k_BT d\, \eta$. The limits of integration
now become $\eta_< = \frac{\mu - \ve}{2 k_BT}$ and $\eta_> = - \infty$. Obviously, $\eta_< > \eta_>$. The integral now takes the form

\begin{equation}
\int\limits_{\eta_<}^{-\infty} \frac{-2 k_BT d\,\eta}{4 k_BT \cosh^2[\eta]} = \frac{1}{2} \int\limits_{-\infty}^{\eta_<} \frac{d\, \eta}{\cosh^2 \eta} = \frac{1}{2} \tanh \eta \vert_{\infty}^{\eta_<} = \frac{1}{2} \left(
\tanh \eta_< + 1 \right) \ .
\end{equation}
Also we can write

\begin{equation}
\tanh \eta_< = \tanh \left[ \frac{\mu - \ve}{2 k_BT} \right] = - \tanh \left[
\frac{\beta}{2} (\ve - \mu)  \right] \ .
\end{equation}
Finally,  we obtain

\begin{equation}
n_F (\xi) = \frac{1}{2} \left( 1 - \tanh\frac{\beta \xi}{2}  \right) =
\int\limits_{0}^{\infty}  d \, \mu' \ \frac{\theta(\mu - \ve)}{4 k_BT \cosh^2 [(\mu - \mu')/(2k_BT)]}
\end{equation}
Alternatively, we have

\begin{equation}
n_F(\ve,\mu; \, T) =\int\limits_{0}^{\infty} \frac{ n_F(\ve,\mu'; \, T=0)
\,\, d \, \mu'}{4 k_BT \cosh^2 [(\mu - \mu')/(2k_BT)]} \, .
\end{equation}

%
%
%

\section{Derivation of the long wavelength limit of the Polarization Function \cite{Wunsch, DS07, pavlo}
and the Finite-temperature Plasmon Modes \cite{SDSLi} }
\label{seca1}

First we need to derive the long-wave limit of the polarization with $q \rightarrow 0$
with fixed frequency $\omega \gg v_F q$. We consider two separate cases with different
imaginary part of the polarization: $1)$ $\omega > 2 \mu$ and $2)$ $\omega < 2 \mu$.
\begin{equation}
 \Pi^{0}(q,\omega) = \frac{g q^2}{8 \pi \hbar \omega}
 \left\{
 \frac{2 \mu}{\hbar \omega} + \frac{1}{2} \ln \left \vert \frac{2 \mu - \hbar \omega}{2 \mu -
 \hbar \omega}\right \vert -\frac{\mathrm{i} \pi}{2}\theta(\hbar \omega - 2 \mu)
 \right\}
 \label{eq11}
\end{equation}
If we consider the region with $\hbar \omega \ll \mu$, we can simplify as follows:
\begin{equation}
 \ln \left \vert \frac{2 \mu - \hbar \omega}{2 \mu - \hbar \omega}\right \vert \backsimeq  - \frac{\
 \hbar \omega}{2 \mu}
\end{equation}
Finally, for the region $1A$ (where the zero-temperature plasmons exist), we have
\begin{eqnarray}
  && {\mbox Re}\  \, \Pi^{0}(q,\omega) = \frac{g q^2}{8 \pi \hbar \omega} \frac{2 \mu}{\hbar \omega} = \frac{\mu}{\pi \hbar^2}
  \frac{q^2}{\omega^2} \\
  \nonumber
  && {\mbox Im}\ \,  \Pi^{0}(q,\omega) = 0, \hspace{0.6 in} (\hbar \omega < 2 \mu)
  \end{eqnarray}
\subsection{Derivation of Eq.(\ref{eq11})}
The general exression for the non-interacting polarization functions is:
\begin{equation}
 \Pi^{0}(q,\omega) = P_0(q,\omega) + \delta P(q,\omega)
\end{equation}
with the term $P_0(q,\omega)$, corresponding to the inter-band transitions, which is urely imaginary for $\omega > v_F q$ and
the $\mu$-dependent $\delta P(q,\omega)$, which appears due to the inrta-band transitions inside the conduction band (for $\mu
> 0$). The terms could be introduced in the following form
\begin{eqnarray}
\label{main1}
 && P_0 (q,\omega) = - \frac{i \pi}{\hbar^2 v_F^2} F_1(q,\omega) \\
 \nonumber
 && \delta P(q,\omega) = - \frac{g \mu}{2 \pi \hbar^2 v_F^2} +\frac{1}{\hbar^2 v_F^2} F_1(q,\omega) \left\{
 F_2\left(\frac{\hbar \omega + 2 \mu}{\hbar v_F q}\right)-F_2\left(\frac{2 \mu- \hbar \omega}{\hbar v_F q}\right)
 - i \pi
 \right\}
\end{eqnarray}
with
\begin{eqnarray}
 && F_1(q,\omega) = \frac{g}{16 \pi} \frac{\hbar v_F^2 q^2}{\sqrt{\omega^2 - v_F^2 q^2}} \\
 \nonumber
 && F_2(\mathcal{X}) =\mathcal{X}\sqrt{\mathcal{X}^2-1} - \ln(\mathcal{X}+\sqrt{\mathcal{X}^2-1})
 \hspace{0.3 in} (\mathcal{X}>1)
\end{eqnarray}
Eq.(\ref{main1}) is satisfied in the \texttt{Region 1A}, given as $\omega < 2 \mu - v_F q$ and $\omega > v_F q$.
Now let us consider each term for $q \rightarrow 0$ and $v_F q \ll \omega$.
We start with the following approximation
\begin{equation}
 \frac{1}{\sqrt{\omega^2 - v_F^2 q^2}} = \frac{1}{\omega} \left( 1 - \left(\frac{v_F q}{\omega}\right)^2  \right)^{-1/2}  \backsimeq
\frac{1}{\omega} + \frac{(v_F q)^2}{2 \omega^3} + \frac{3}{8}\frac{v_F^4 q^4}{\omega^5}
 \end{equation}
Consequently, the inter-band polarization function has the form
\begin{equation}
 P_0 (q,\omega) = - \frac{i \pi}{\hbar} \frac{q^2}{\omega} - \frac{i \pi}{\hbar} \frac{v_F^2 q^4}{2 \omega^3} + ... \, ,
\end{equation}
i.e. is purely imaginary.
\par
Now we consider the intra-band part $\delta P(q,\omega)$ . First, we analyze how each term of it behaves
for $q \rightarrow 0$. Thus
\begin{equation}
\mathcal{X} = \frac{2 \mu \pm \hbar \omega}{\hbar v_F q} \gg 1
\end{equation}
and
\begin{eqnarray}
&& F_2(\mathcal X) = \mc{X}\sqrt{\mc{X}^2 - 1} - \ln[\mc{X} + \sqrt{\mc{X}^2 -1}], \hspace{1 in} (\mc{X} \gg 1) \\
\nonumber
&& \mc{X}\sqrt{\mc{X}^2 - 1} = \mc{X}^2 (1 - 1/\mc{X}^2)^{1/2} \backsimeq  \mc{X}^2 - \frac{1}{2} \\
\nonumber
&& \mc{X} - \sqrt{\mc{X}^2 - 1} = \mc{X}^2 - \mc{X} \left(1 - \frac{1}{2 \mc{X}^2 }   \right) \backsimeq \mc{X}^2 - \mc{X}
\end{eqnarray}
Let us summarize the results:
\begin{eqnarray}
\nonumber
&& F_2 \left( \frac{\hbar\omega + 2 \mu }{\hbar v_F q} \right) - F_2 \left( \frac{\hbar\omega - 2 \mu }{\hbar v_F q}
 \right) \backsimeq \left( \frac{\hbar\omega + 2 \mu }{\hbar v_F q} \right)^2 - \frac{1}{2} - \left( \frac{\hbar\omega -
 2 \mu }{\hbar v_F q} \right)^2 + \frac{1}{2} - \ln\left[ \left( \frac{\hbar\omega + 2 \mu }{\hbar v_F q} \right)^2
 \right] + \ln\left[  \left( \frac{\hbar\omega - 2 \mu }{\hbar v_F q} \right)^2
 \right]  = \\
&& = \frac{2\times 4 \hbar \omega \mu }{(\hbar v_F q)^2} + 2 \ln\left[ \frac{2 \mu - \hbar \omega}{2 \mu + \hbar \omega} \right]
\end{eqnarray}
and also
\begin{equation}
 \frac{1}{\hbar^2 v^2} F_1 (q,\omega) \backsimeq \frac{g}{16 \pi \hbar} \frac{q^2}{\omega} + \frac{g}{32 \pi \hbar}
 \frac{v_F^2 q^4}{\omega^2}
\end{equation}
As a result
\begin{equation}
\frac{1}{\hbar^2 v^2} F_1 (q,\omega) \left[
F_2 \left( \frac{\hbar\omega + 2 \mu }{\hbar v_F q} \right) - F_2 \left( \frac{\hbar\omega - 2 \mu }{\hbar v_F q}
 \right)
\right] = \frac{\mu g}{2 \pi \hbar^2 v_F^2} + \frac{g q^2}{4 \hbar^2} \frac{\mu}{\pi \omega^2}
\end{equation}
Finally, we can write
\begin{equation}
{\mbox Re}\  \delta P(q,\omega) = - \frac{g \mu}{2 \pi \hbar^2 v_F^2} + \frac{\mu g}{2 \pi \hbar^2 v_F^2} +
\frac{g q^2}{8 \pi \hbar \omega} \left( \frac{2 \mu}{\hbar \omega} \right) + \frac{g q^2}{16 \pi
\hbar \omega}  \ln\left[ \frac{2 \mu - \hbar \omega}{2 \mu + \hbar \omega} \right]
\end{equation}
As far as the imaginary part is concerned, it is equal to zero everywhere in \texttt{Region 1A}:
\begin{equation}
{\mbox Im}\ \delta P(q, \omega) = i \pi \frac{F_1(q,\omega)}{\hbar^2 v_F^2}
\end{equation}
According to Eq.(\ref{main1})
\begin{equation}
P_0 (q,\omega) = - \frac{i \pi}{\hbar^2 v_F^2} F_1(q,\omega)
\end{equation}
so that
\begin{equation}
\Pi^0 (q,\omega) = P^0(q, \omega) + \delta P(q, \omega) = 0
\end{equation}
So Eq.(\ref{eq11}) is confirmed.
\subsection{Finite-temperature plasmons}
The finite temperature non-interaction polarization function is expressed as follows:
	\begin{equation}
	\Pi^{(0)}_T(q,\omega) = \int\limits_{0}^{\infty} \frac{\Pi^{0}(q,\omega)}{4 k_BT \left( \cosh \left[ \frac{\mu - \mu'}{2 k_BT}
		\right] \right)^2}
	\end{equation}
	In the long-wave limit approximation, the zero-temperature polarizability is as follows:
		\begin{equation}
		\Pi^0 (q, \omega) = \frac{g q^2}{4 \pi} \frac{\mu}{(\hbar \omega)^2} - i \frac{g q^2}{16
		\hbar \omega} \theta(\hbar \omega - 2 \mu)
		\end{equation}
Let us first consider the real part:
\begin{equation}
{\mbox Re}\  \, \Pi^{(0)}_T = \int\limits_{0}^{\infty} \frac{g q^2}{4 \pi} \frac{\mu'}{(\hbar \omega)^2}
\frac{1}{4 k_BT} \frac{d\, \mu'}{\cosh^2 \left[(\mu - \mu')/(2 k_BT)\right]} = \frac{g q^2}{4 \pi
\hbar^2 \omega^2} \frac{1}{4 k_BT} \int\limits_0^{\infty} \frac{\mu' \,\, d\, \mu' }{\cosh^2
\left[ \frac{\mu - \mu'}{2 k_BT} \right]}
\end{equation}
This integral could be easily evaluated:
\begin{equation}
\int\limits_0^{\infty} \frac{\mu' \,\, d\, \mu' }{\cosh^2 \left[ \frac{\mu -
\mu'}{2 k_BT} \right]} = 4 k^2 T^2 \ln \left[ 1 + \tet{e}^{\beta \mu}\right]
\end{equation}
We consider \textit{intrinsic} graphene with $\mu = 0$ so that $\ln \left[ 1 + \tet{e}^{\beta \mu}\right] = \ln 2 $.
\par
Thus
\begin{equation}
{\mbox Re}\  \, \Pi^{(0)}_T (q, \omega )= \frac{g \ln 2}{2 \pi} \frac{q^2}{\hbar^2 \omega^2} k_BT
\end{equation}
\par
The imaginary part is as following:
\begin{equation}
{\mbox Im}\ \, \Pi^{(0)}_T (q, \omega) = - i \frac{g q^2}{16 \hbar \omega} \int\limits_{0}^{\infty}
\frac{d\,\mu'}{4 k_BT} \frac{\theta(\hbar \omega - 2 \mu' )}{\cosh^2 \left[ \frac{\mu - \mu'}{2 k_BT} \right]}
\end{equation}
Finally
\begin{equation}
{\mbox Im}\ \, \Pi^{(0)}_T (q, \omega) = i \frac{g q^2}{16 \hbar \omega} \frac{1}{2} \tanh
\frac{\hbar \omega}{4 k_BT} \backsimeq i \frac{g}{128} \frac{q^2}{k_BT}
\end{equation}
\subsection{plasmons}
Now we rewrite the finite-temperature polarization function:
\begin{equation}
\Pi^{(0)}_T (q,\omega) = \frac{2 \ln 2}{\pi} \frac{q^2}{\hbar^2 \omega^2} k_BT + \frac{i}{16} \frac{q^2}{k_BT}
\label{pol}
\end{equation}
The dielectric function $\epsilon (q, \omega)$ is
\begin{equation}
\epsilon (q, \omega) = 1 - V(q) \Pi^{(0)}_T (q, \omega) \,
\end{equation}
where $V(q) = 2 \pi e^2 / (\epsilon_s q)$.
\par
We introduce real and imaginary parts of the frequency $\omega \rightarrow \omega + i \gamma $ and $r_s = e^2/(\epsilon_s \hbar v_F)$, we find the real part:
\begin{equation}
1 - \frac{2 \pi e^2}{\epsilon_s q} \frac{2 \ln 2}{\pi} \frac{q^2}{\hbar^2 \omega^2} k_BT = 0
\end{equation}
we obtain:
\begin{equation}
\label{plasmon}
\omega^2 = \frac{4 \ln 2}{\hbar} v_F  \, r_S \, q \, k_BT
\end{equation}

\section{Imaginary part of the frequency}
For $\Delta = 0$ the polarization function is
\begin{equation}
\Pi^{(0)}_T (q,\omega) = \frac{2 \ln 2}{\pi} \frac{q^2}{\hbar^2 \omega^2} k_BT + \frac{i}{16} \frac{q^2}{k_BT}
\end{equation}
The plasmons are defined by the following equation
\begin{equation}
\epsilon (q, \omega) = 1 - \frac{2 \pi e^2}{\epsilon_s q} \Pi^{(0)}_T (q, \omega) = 1 - \frac{2 \pi}{q}  r_s \hbar v_F  \Pi^{(0)}_T (q, \omega) = 0\,
\end{equation}
Let us introduced the complex frequency:
\begin{equation}
\omega \Longrightarrow \omega + i \gamma
\end{equation}
\begin{equation}
\frac{1}{\omega^2} \Longrightarrow \frac{1}{(\omega + i \gamma)^2} \backsimeq \frac{1 - 2 i \gamma/ \omega}{\omega^2} =\frac{1}{\omega^2} - 2 i \frac{\gamma}{\omega^3}
\end{equation}
This results in the following equation
\begin{equation}
1 - \frac{2 \pi}{q} r_s \hbar v_F \left\{
\frac{2 \ln 2}{\pi \hbar^2} q^2 k_BT \left(
\frac{1}{\omega^2}- \frac{2 i \gamma}{\omega^3}
\right) +
\frac{i}{16} \frac{q^2}{k_BT}
\right\} = 0
\end{equation}
According to Eq.(\ref{plasmon}),
\begin{equation}
1 - \frac{2 \pi}{q} r_s \hbar v_F \left\{
\frac{2 \ln 2}{\pi \hbar^2} q^2 k_BT
\frac{1}{\omega^2} \right\} = 0
\end{equation}
which leads to
\begin{equation}
-\frac{2 \ln 2}{\pi \hbar^2} q^2 k_BT \frac{2 i \gamma}{\omega^3} + \frac{i}{16} \frac{q^2}{k_BT} = 0
\end{equation}
and
\begin{equation}
\gamma = \frac{\pi \hbar^2}{\ln 2} \, \frac{\omega^3}{64 k^2 T^2}
\end{equation}
\begin{equation}
\omega^3 = 8 \hbar^{-3/2} \left( \ln 2 \right)^{3/2} v_F^{3/2} r_S^{3/2} q^{3/2} (k_BT)^{3/2}
\end{equation}
\begin{equation}
\gamma = \frac{\pi}{8} \frac{\hbar^{1/2}}{\sqrt{k_BT}} \left( v_F r_s q \right)^{3/2}
\end{equation}

\end{document}